\documentclass[aps,prl,twocolumn,nofootinbib,longbibliography,notitlepage]{revtex4-1}
\usepackage{graphicx}
\usepackage{amssymb}
\usepackage{hyperref}
\usepackage{amsmath}
\usepackage{amsfonts}
\usepackage{subfigure}
\usepackage{dsfont}
\usepackage{times}
\usepackage{dcolumn}
\usepackage{esint}
\usepackage{mathrsfs}
\usepackage{bm}
\usepackage{multirow}
\usepackage{color}
\usepackage{datetime}
\usepackage{braket}
\usepackage{romannum}

\begin{document}
\title{Detecting bulk and edge exceptional points in non-Hermitian systems through  generalized Petermann factors}
\author{Yue-Yu Zou}
\thanks{These authors contributed equally to this work.}
\author{Yao Zhou}
\thanks{These authors contributed equally to this work.}
\author{Li-Mei Chen}
\thanks{These authors contributed equally to this work.}
\author{Peng Ye}
\email{yepeng5@mail.sysu.edu.cn}
\affiliation{School of Physics, State Key Laboratory of Optoelectronic Materials and Technologies, and Guangdong Provincial Key Laboratory of Magnetoelectric Physics and Devices, Sun Yat-sen University, Guangzhou, 510275, China}
\date{\today}
\begin{abstract}
Non-orthogonality in  non-Hermitian quantum systems gives rise to tremendous exotic quantum phenomena, which can be fundamentally traced back to  non-unitarity and is much more fundamental and universal than complex energy spectrum. In this paper, we introduce an interesting quantity (denoted as $\eta$) as a new variant of the Petermann factor to directly and efficiently measure   non-unitarity and the associated  non-Hermitian physics.
By tuning the  model parameters of   underlying non-Hermitian systems, we find that the discontinuity  of both $\eta$ and its first-order derivative (denoted as $\partial \eta$)  pronouncedly  captures rich   physics that is fundamentally caused by non-unitarity. More concretely, in the 1D non-Hermitian topological systems, two mutually orthogonal edge states that are respectively  localized on two boundaries become non-orthogonal in the vicinity of discontinuity of $\eta$ as a function of the model parameter, which is dubbed  ``{edge state transition}''. Through theoretical analysis, we identify that the appearance of  edge state transition indicates the existence of exceptional points~(EPs) in topological edge states. Regarding the discontinuity of $\partial\eta$, we investigate a two-level non-Hermitian model and establish a connection between the points of discontinuity of $\partial \eta$ and EPs of bulk states. By studying this connection in more general lattice models, we find  that some models  have discontinuity of $\partial\eta$, implying the existence of EPs in bulk states.
\end{abstract}
\maketitle

{\color{blue}\emph{Introduction.}}---
Recently, non-Hermitian systems~\cite{benderMaking2007,caoDielectric2015,rotterNonHermitian2009,ashidaNonHermitian2020,bergholtzExceptional2021} have drawn  great interests, due to exotic quantum phenomena, e.g., generalized bulk-edge correspondence~\cite{yaoEdge2018,yokomizoNonBloch2019,yangNonHermitian2020}, exceptional points~(EPs)~\cite{heissPhysics2012,gaoObservation2015,zhenSpawning2015,hahnObservation2016,zhangObservation2017,miriExceptional2019,wangArbitrary2019,xiaoObservation2021,bergholtzExceptional2021,huKnot2022}, non-Hermitian skin effect~\cite{yaoEdge2018,kunstBiorthogonal2018,leeAnatomy2019,okumaTopological2020,okumaNonHermitian2021,zhangUniversal2022} and unidirectional invisibility~\cite{linUnidirectional2011}. To gain a theoretically understanding of these phenomena, non-Bloch band theory~\cite{yaoEdge2018, yaoNonHermitian2018,yokomizoNonBloch2019,dengNonBloch2019,liuSecondOrder2019,longhiNonBlochBand2020,yangNonHermitian2020,zhangCorrespondence2020,kawabataNonBloch2020}, which generalizes the conception of Brillouin zone, is established and applied to analyze the topological phase~\cite{shenTopological2018,bergholtzExceptional2021} in non-Hermitian systems. Remarkably, non-Hermitian skin effect, the natural consequence of generalized Brillouin zone, is found to have  a connection to EPs~\cite{zhangUniversal2022}. Meanwhile, compared with Hermitian systems, the classification~\cite{gongTopological2018,kawabataClassification2019,kawabataSymmetry2019,zhouPeriodic2019,herviouEntanglement2019,wojcikHomotopy2020} of topological phases in non-Hermitian systems has been significantly enriched.  In addition, from the quantum-informative perspective, quantum entanglement properties of non-Hermitian systems~\cite{changEntanglement2020,chenEntanglement2021,sayyadEntanglement2021,guoEntanglement2021,chenQuantum2022} display highly unusual features in entanglement entropy and entanglement spectrum. Besides the crystalline system mentioned before, non-Hermiticity has also been introduced to noncrystalline systems, e.g. quasi-crystal systems~\cite{longhiTopological2019,zengWinding2020,liuNonHermitian2020,liuLocalization2021, chenQuantum2022} and disorder systems~\cite{hatanoLocalization1996,hatanoVortex1997,hatanoNonHermitian1998,linObservation2022}.

While complex energy spectrum is commonly observed in many non-Hermitian systems, real energy spectrum generally holds in systems with  $\mathcal{PT}$ symmetry~\cite{benderIntroduction2005,benderMaking2007}. In our prior work~\cite{chenQuantum2022}, a non-Hermitian quasi-crystal model alway has complex energy spectrum which can not be used to characterize its the phase transition.  Thus, compared with non-unitarity, it is apparently that complex energy spectrum is not a special sign of non-Hermitian physics. In this paper, to explore physics of non-Hermitian systems, we focus the property of non-unitarity, i.e., the non-orthogonal eigenvectors~\cite{wiersigNonorthogonality2019} which induce interesting phenomena in various research areas~\cite{petermann1979calculated,makris2008beam,schomerus2009excess,wiersig2011nonorthogonal,fyodorov2012statistics,makris2014anomalous,davy2018selectively,davy2019probing}.   When eigenvectors are not mutually orthogonal,   the familiar inner product in  Hermitian systems is no longer valid and the usual definition of quantum expectation of operators is no longer proper.  To proceed further, in the literature, the idea of bi-orthogonal basis is introduced. More concretely, for a non-Hermitian Hamiltonian $H$, the   right eigenvectors $|R,n\rangle$ obey $H|R,n\rangle =E_n|R,n\rangle$, and left eigenvectors $|L,n\rangle$ obey $H^\dagger |L,n\rangle =E_n^*|L,n\rangle$, then bi-orthogonality relation can be represented as $\langle L,n|R,m\rangle =\delta_{n m}$. When the system remains unitarity, $\langle L,n|L,m\rangle =\langle R,n|R,m\rangle =\delta_{n m}$. Thanks to the bi-orthogonality relation, many theoretical approaches originally introduced in Hermitian systems can be borrowed to study non-Hermitian systems. Therefore, a series of physical conceptions are reproduced in non-Hermitian systems~\cite{shenTopological2018,kunstBiorthogonal2018, yaoNonHermitian2018,songNonHermitian2019}.

By using bi-orthogonal basis, we can   study non-Hermitian quantum systems by constructing Hilbert space. However, it is still unclear how to simply and efficiently characterize non-unitarity arising from the non-orthogonality among right-eigenvectors (or left-eigenvectors). To discuss non-unitarity of non-Hermitian systems, without loss of generality, we focus on studying the property of right basis but not the whole bi-orthogonal basis in this work~(because left eigenvectors have the similar property with right eigenvectors). In this paper, generalizing the idea of Lee-Wolfenstein bound~\cite{leeAnalysis1965,wiersigNonorthogonality2019} for all eigenstates of a non-Hermitian system,  we define a quantity to measure the strength of non-unitarity of non-Hermitian systems as follows:
\begin{align}
    \eta = \frac{\sum_{n<m}|\langle R,n|R,m\rangle|^2}{\sum_{n<m}|\langle R,n|R,n\rangle||\langle R,m|R,m\rangle|}\,,\label{definition_eta}
\end{align}
where $0\le \eta\le 1$. When  $\eta=0$, the system is unitary with mutually orthogonal eigenvectors. On the contrary, when $\eta=1$, the eigenvectors are totally coalescent, resulting in the extreme case of non-unitarity. Additionally, the definition of the quantity $\eta$ can be considered as a new variant of the Petermann factor which has various definitions as given in Refs.~\cite{ashidaNonHermitian2020,petermann1979calculated,wangPetermannfactor2020}.

\begin{table}[ht]
  \caption{The correspondence between the location of EPs and the discontinuity of $\eta$ and its derivative $\partial\eta$ in the Mode I-IV.}
\label{tab:table}
\begin{ruledtabular}
\begin{tabular}{ccc}
 \multicolumn{3}{c}{The correspondence between $\eta$, $\partial\eta$ and EPs}       \\  \colrule
  Discontinuity &  $\eta$     &   $\partial\eta$          \\
  EPs &  Topological edge states    &         Bulk states                            \\
   Model  &   I       &    II, III, IV
\end{tabular}
\end{ruledtabular}
\end{table}

In this paper, we study the behavior of the quantity $\eta$ in various interesting non-Hermitian models as the system parameters vary. We observe the various behaviors of the quantity $\eta$, such as the discontinuity of the quantity $\eta$ and its first-order derivative $\partial\eta$, which imply respectively  the existence of EPs in topological edge and bulk states of non-Hermitian systems, as illustrated in Tab.~\ref{tab:table}. Specifically, when a non-Hermitian topological systems undergoes an \textit{edge state transition}, where the orthogonal edge states become non-orthogonal, the quantity $\eta$ would have discontinuity point which imply EPs appearing in the topological edge states. For studying the physical consequence causing the discontinuity of $\partial\eta$, we utilize a two-level model exhibiting that when the quantity $\eta$ near the EP of bulk states, $\partial\eta$ would become discontinuous. Thus, this feature of $\eta$ can be considered evidence for identifying the existence of EPs in bulk states. Furthermore, using this feature, we infer that the bulk states of some non-Hermitian lattice systems possess EPs.



\begin{figure}
        \centering
        \includegraphics[width=8.9cm]{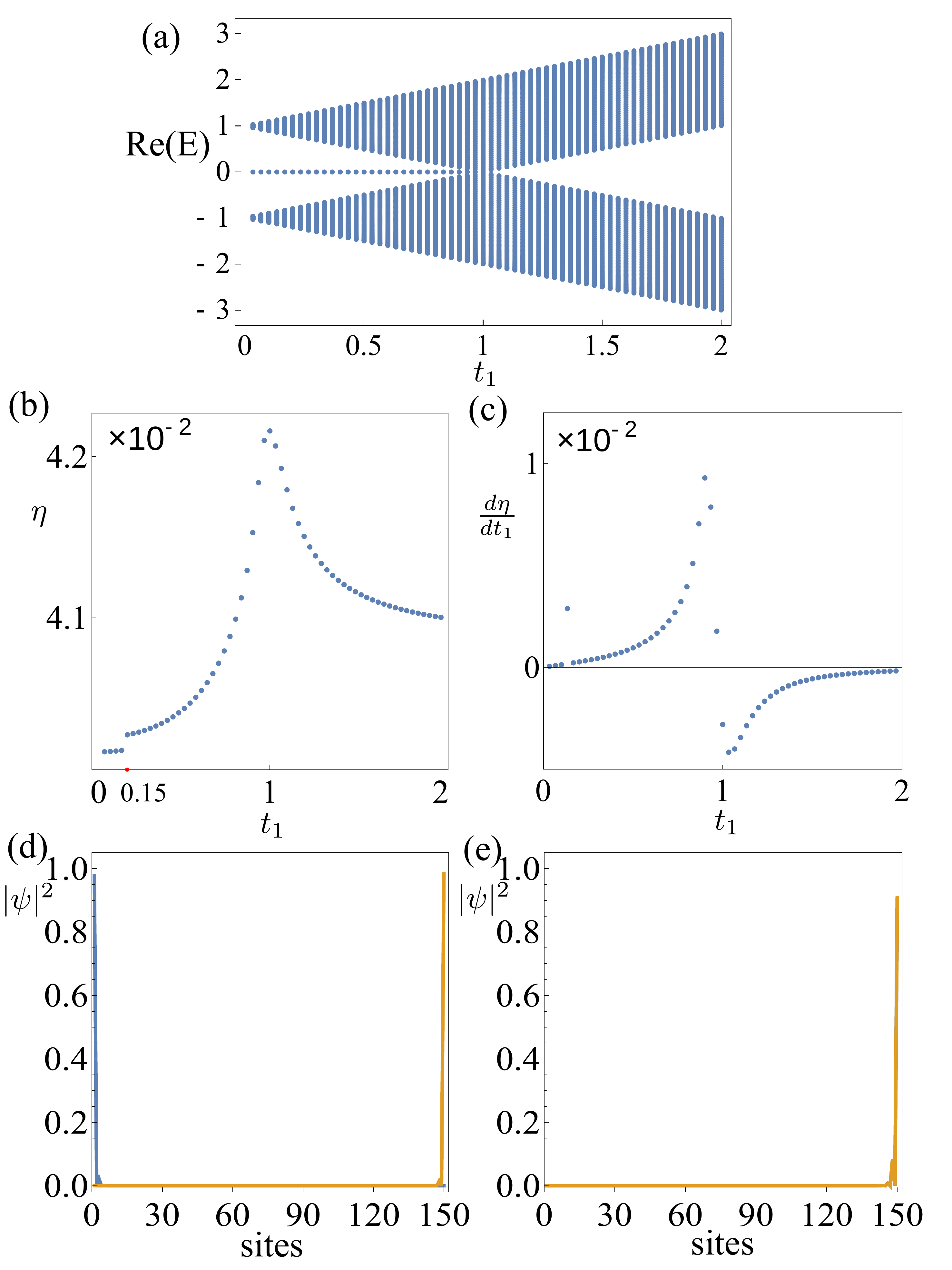}
        \caption{(a) Real part of energy spectrum of the model-I~\eqref{ssh_model} as a function of $t_1$. The topological phase transition occurs at $t_1=\sqrt{t_2^2-g^2}\approx 0.99$. (b) and (c) respectively show the quantity $\eta$ and its derivative as a function of $t_1$. The discontinuity of $\eta$ and the local maximum of $\eta$ appear respectively at $t_1\approx 0.15$ and $t_1\approx 0.99$. (d) and (e) respectively demonstrate two distributions of edge states at $t_{1}=0.133$ and $t_{1}=0.333$ near the discontinuity point $t_{1}=0.15$ of the quantity $\eta$. Here, $t_2=1, g=0.1$, the length of the system~\eqref{ssh_model} $L=150$.}
        \label{sshmodel}
\end{figure}

{\color{blue}\emph{Edge state transition and edge EPs at the discontinuity of $\eta$.}}---
To focus on the nature of non-unitarity of non-Hermitian systems, we consider the behavior of the quantity $\eta$ in a $1$D non-reciprocal Su-Schrieffer-Heeger~(SSH) model~\cite{chengCompetition2022}:
\begin{equation}
\begin{aligned}
  \textbf{Model-I} ~H = &\sum_n [t_1 c_{n,A}^\dagger c_{n,B}+t_1 c_{n,B}^\dagger c_{n,A}+\\&(t_2+g) c_{n,B}^\dagger c_{n+1,A}+(t_2-g) c_{n+1,A}^\dagger c_{n,B} ],
\end{aligned}
\label{ssh_model}
\end{equation}
where $c_{n,A}$ ($c_{n,B}$)  respectively denote annihilation operators of spinless fermions at sublattice A (B) in the $n$th unit cell. We restrict the parameters $g$, $t_{1,2}$ in the real regime. When the parameters satisfy the condition $|t_1|<\sqrt{t_2^2-g^2}$, the  system is in a non-Hermitian topological phase with non-trivial winding number and two topological edge states. When $|t_1|>\sqrt{t_2^2-g^2}$, the system is in a trivial phase without topological edge states. Thus, a  topological phase transition occurs at   $|t_1|=\sqrt{t_2^2-g^2}$, which can be identified by the appearance/disappearance of zero energy modes in Fig.~\ref{sshmodel}(a). Meanwhile, we study the quantity $\eta$ as a function of $t_{1}$ in the model-I~\eqref{ssh_model}. We find that the phase transition point coincides with the local maximum of $\eta$ in Fig.~\ref{sshmodel}(b), while the derivative $\partial\eta$ in Fig.~\ref{ssh_model}(c) is continuous at the transition point. Based on the method of generalized Brillouin zone in Ref.~\cite{yaoEdge2018,yokomizoNonBloch2019,yangNonHermitian2020}, we analytically obtain the effective bulk Hamiltonian of the model-I~\eqref{ssh_model} with open boundary condition (more detailed derivation see in supplemental materials (SM) part-A). By using the effective bulk Hamiltonian, when $t_{2}=1$ and $g=0.1$, the bulk states at topological transition point of the model-I~\eqref{ssh_model} do not have EPs.

Furthermore,  we realize that this model exhibits a significant discontinuity of $\eta$   at  $t_1=0.15$ in Fig.~\ref{sshmodel}(b).  To clarify the appearance of the discontinuity of $\eta$, we plot the edge states of the system of two parameter points $t_{1} = 0.133$ and $t_{1}=0.333$ near the discontinuity point $t_{1}=0.15$ respectively in Fig.~\ref{sshmodel}(d) and (e). We find that the topological edge states separately localized at two boundaries are orthogonal in Fig.~\ref{sshmodel}(d), while in Fig.~\ref{sshmodel}(e), the  two edge states are simultaneously localized at one boundary and become non-orthogonal. This phenomena are found in recent Refs.~\cite{wangDefective2020,chengCompetition2022}, and we successfully connect the phenomena to the discontinuity of $\eta$.

Next, we analytically obtain the topological edge states of the model-I~\eqref{ssh_model} in topological phase to explain the appearance of discontinuity points of $\eta$. As discussed in SM part-B, when consider the thermodynamic limit ($N\rightarrow \infty$), the two zero-energy edge states are expressed as
$\phi_{n,A} = (-\frac{t_{1}}{t_{2}-g})^{n-1}\phi_{1,A}$ and  $\phi_{1,B} = (-\frac{t_{1}}{t_{2}+g})^{n-1}\phi_{n,B}$, where $\phi_{n,A(B)}$ is the wavefunction on the sublattice $A(B)$ in the $n$th unit cell. From the expression of $\phi_{n,A(B)}$, we can determine the localization behaviors of the edge states. Furthermore, to satisfy the boundary conditions $\phi_{1,B}=\phi_{N,A}=0$ ( here set $t_{2}=1,g=0.1$), when $t_{1}<t_{2}-g$, the wavefunctions $\phi_{n,A}$ and $\phi_{n,B}$ are respectively localized at the left and right endpoints of the 1D chain, and have no contribution to $\eta$. When  $t_{1}>t_{2}-g$,  to satisfy the boundary condition, we find the wavefunction $\phi_{n,A}$ should satisfy the relation $\phi_{n,A}=\phi_{N,A}=0$. Then, the wavefunction $\phi_{n,A}$ disappears. For this reason,  we can consider the two topological edge states merge into one topological edge state and are simultaneously localized at the right endpoint of the 1D chain and have contribution to $\eta$. Therefore, based on above discussion, we propose that the model-I~\eqref{ssh_model} has edge state transition which causes the discontinuity of $\eta$ with the parameter $t_{1}$ varying and satisfy our numerical results in Fig.~\ref{sshmodel}(d) and (e). Furthermore, based on Ref~\cite{wangDefective2020}, we find that the edge state transition of this model-I~\eqref{ssh_model} is induced by EPs in topological edge states, where the EPs are called edge EPs.  Moreover, it should be noted that the numerical precision of diagonalizing the Hamiltonian matrix of the model-I~\eqref{ssh_model} would influence the location of discontinuity points of $\eta$, where this phenomenon originates from the finite-size effect.

\begin{figure}[htbp]
\centering
\includegraphics[width=8.7cm]{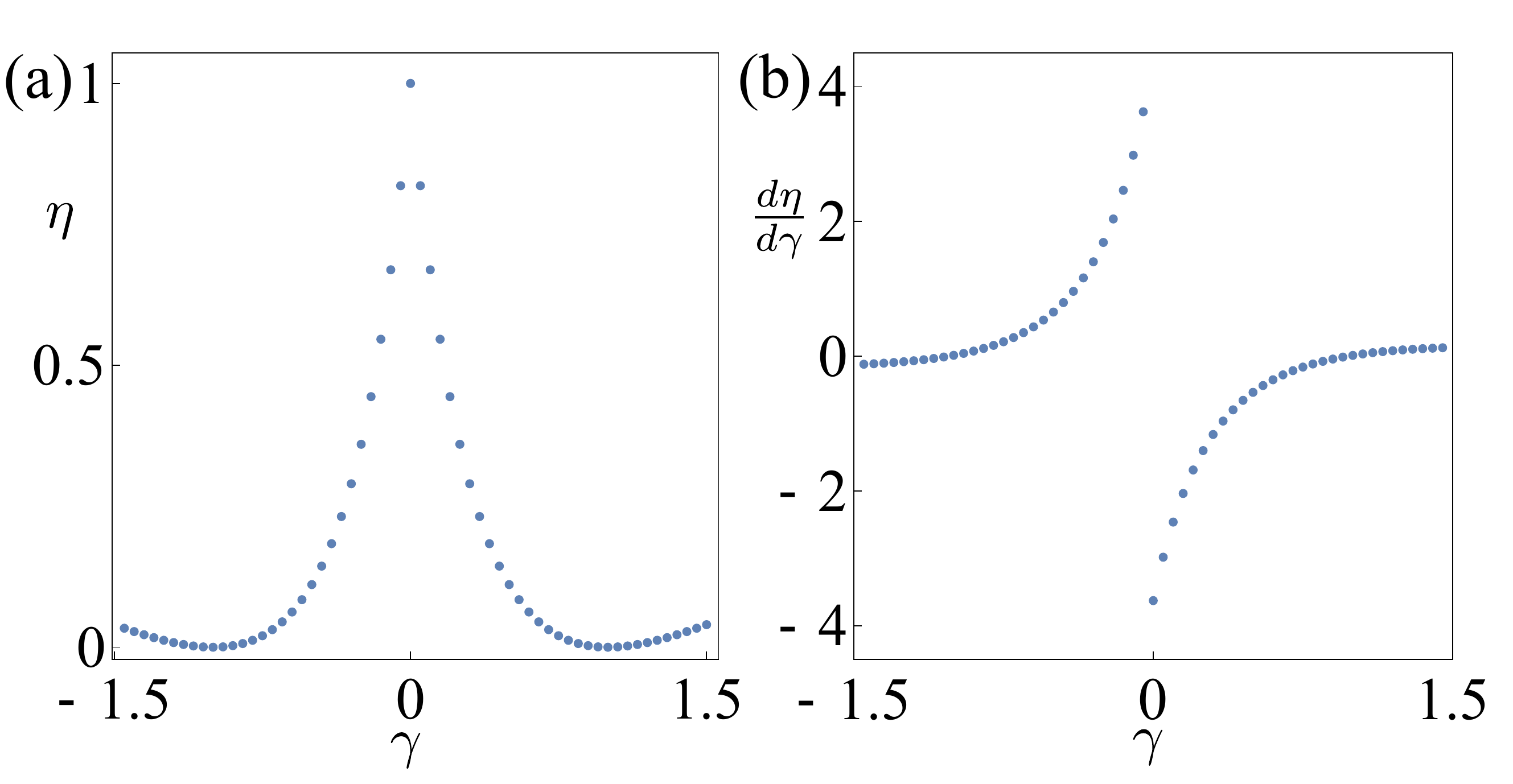}
\caption{(a) and (b) respectively represent the quantity $\eta$ and its derivative as a function of the parameter $\gamma$ in the two-level model (\ref{two_level}).}
\label{twolevel_model}
\end{figure}

{\color{blue}\emph{Bulk EP at the discontinuity of $\partial\eta$.}}---
 In the following, we move to the physics of discontinuity of $\partial\eta$, i.e., the first order derivative of $\eta$. For the purpose, as a warm-up, we first introduce a two-level system to study the behavior of the quantity $\eta$:
\begin{align}
  \textbf{Model-II} ~~~~  H_0 =\begin{pmatrix}
    0 & \gamma\\
    1 & 0
    \end{pmatrix}\,,
    \label{two_level}
\end{align}
where $\gamma\in\mathbb{R}$. By diagonalization, the (right-)eigenvectors of $H_0$  can be  obtained and written as $(\pm\sqrt{\gamma},1)^T$, which results in an analytic form of $\eta=\frac{|1-|\gamma||^2}{(1+|\gamma|)^2}$. It is apparent that  when $\gamma=1$, the model $H_0$ becomes Hermitian with  $\eta=0$. On the contrary, when $\gamma=0$,   $H_0$ reduces to a lower triangular matrix which describes a typical EP, and the quantity $\eta=1$. Next, we study the behavior of the quantity $\eta$ as a function of $\gamma$ near the EP. When  $\lambda=\pm\delta$ and $\delta\rightarrow 0^{+}$, the quantity $\eta|_{\lambda = \delta^{\pm}}=\frac{(1 \mp\delta)^{2}}{(1 \pm \delta)^{2}}|_{\delta\rightarrow 0^{+}} = 1$, and the derivative $\frac{\partial\eta}{\partial\delta}|_{\lambda\rightarrow 0^{\pm}}=\mp 4$. As shown in Fig.~\ref{twolevel_model}, we find that   $\eta$ at EP has a peak and its derivative (denoted as $\partial\eta$) is discontinuous, which is satisfied with our discussion. In the following, we will show that this feature of $\eta$ can be regarded as an evidence to identify EPs of bulk states in more general non-Hermitian quantum systems.

Since this model-II~\eqref{two_level} has merely two levels, the quantity $\eta$ at EP can take the maximum value $1$ and all eigenvectors are coalescent. However, for  models with more than two levels, it usually has various EPs with different degeneracies. Consequently, the eigenvectors are not totally coalescent, and the upper bound (denoted as $\eta_c$) of $\eta$ depends on the configuration of EPs: $\eta_c = \frac{\sum_{n}d_{n}(d_{n}-1)}{N(N-1)}\leq 1$,
where $N = \sum_{n}d_{n}$ is the dimension of Hamiltonian matrix of non-Hermitian systems and $n$  represents $n$th Jordan block with $d_{n}$-fold degeneracy~\cite{zhouyao}. Only when the non-Hermitian system has one EP with $N$-degeneracy, the quantity $\eta_c$ equal to $1$~\cite{zhouyao}.

\begin{figure}[htbp]
\centering
\includegraphics[width=8.8cm]{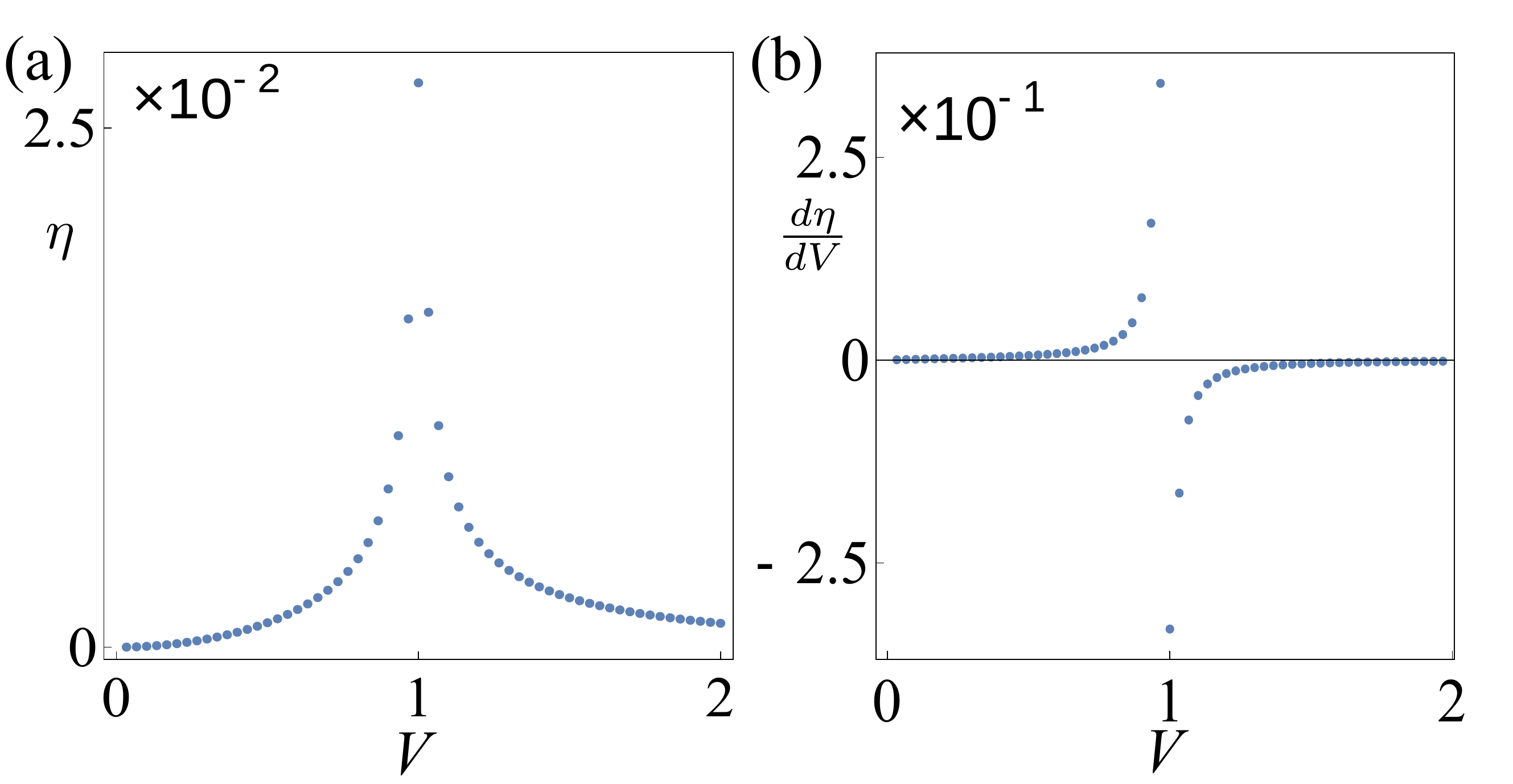}
\caption{(a) and (b) respectively show the quantity $\eta$ and its derivative as a function of the potential strength $V$ in the model-III (\ref{quasi_model}) \cite{chenQuantum2022}. Here, $J_R=1, J_L=0.5$.}
\label{quasimodel}
\end{figure}

To illustrate the physics of the discontinuity of $\partial\eta$, we will study two concrete non-Hermitian lattice models. Firstly, we consider a non-Hermitian quasi-crystal lattice model~\cite{chenQuantum2022} which has a localization-delocalization transition induced by non-Hermiticity:
\begin{equation}
  \textbf{Model-III} ~~ H = \sum_n (J_R c_{n+1}^\dagger c_n +J_L c_{n}^\dagger c_{n+1})+\sum_n V_n c_n^\dagger c_n,
    \label{quasi_model}
\end{equation}
where $c_n(c_n^\dagger)$ is the annihilation (creation) operator of spinless fermion at the $n$th lattice site. $V_n= V \exp(-2\pi i \alpha n)$ is a site-dependent incommensurate complex potential parameterized by an irrational number $\alpha$. The potential strength $V$ is positive and real. We set the parameter $\alpha=\sqrt{2} \approx \frac{239}{169}$ same as Ref.~\cite{chenQuantum2022}. In the practical simulations, we set  the length $L=169$ of the system with periodic boundary condition. As discussed in Ref.~\cite{chenQuantum2022}, metal-insulator phase transition (MIT) of this model-III~\eqref{quasi_model}   occurs at the point $V=1$. In Fig.~\ref{quasimodel}, we can see that the quantity $\eta$ as a function of $V$ exhibits a sharp peak   at  $V=1$, and a discontinuity point of the derivative of $\eta$ coincides with $V=1$ point. These features of $\eta$ in the model-III~\eqref{quasi_model} are similar with the features in the two-level   model~\eqref{two_level}. Therefore, we infer that the model-III~\eqref{quasi_model} have EPs of bulk states at the MIT transition point, where the EPs are called bulk EPs.


\begin{figure}[htbp]
  \centering
  \includegraphics[width=8.8cm]{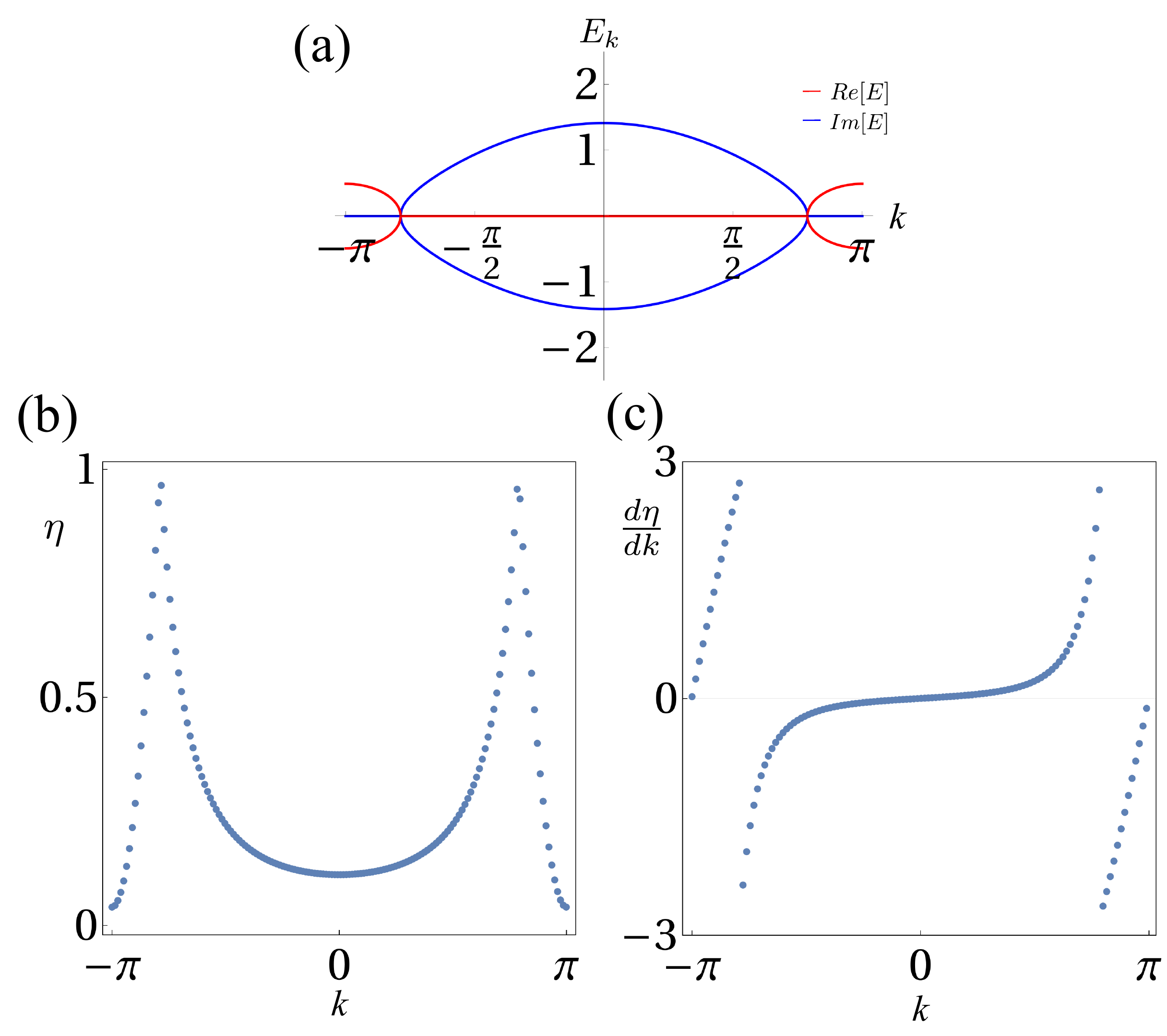}
  \caption{(a) is the energy spectrum of the Hamiltonian~\eqref{ptsshmodel}, (b) and (c) are respectively the quantity $\eta$ and its derivative with the wave vector $k$ varying. Here $(w,u,v)=(0.7,0.5,0.8)$. }
  \label{ptssheta}
  \end{figure}

Furthermore, we study a $\mathcal{PT}$-symmetrical SSH model~\cite{changEntanglement2020} with EPs in bulk states to show the behaviors of the quantity $\eta$, where the model is written in momentum space as
\begin{equation}
  \textbf{Model-IV} ~~ H(k) =
  \begin{bmatrix}
    i u & w e^{-i k} +v \\
    w e^{i k} +v & -i u
  \end{bmatrix},
  \label{ptsshmodel}
\end{equation}
where the parameters $u, w ,v \in \mathbb{R}$, and $k$ is the wave vector (or momentum). $\mathcal{PT}$ symmetry is represented as $\sigma_{x}H(k)\sigma_{x}=H^{*}(k)$. The energy dispersion of the model~\eqref{ptsshmodel} is $ E(k)=\pm\sqrt{|w e^{- i k}+v|^{2}-u^{2}}$. Due to absence of skin effect~\cite{yangNonHermitian2020}, we can use the eigenstates of the Hamiltonian~\eqref{ptsshmodel} faithfully representing the bulk states of $\mathcal{PT}$-symmetrical SSH model with open boundary condition. Meanwhile, as discussed in ~\cite{changEntanglement2020}, with $u, w, v >0$, the $\mathcal{PT}$-broken phase $(|w-v|<u )$ of the model-IV~\eqref{ptsshmodel} has complex energy spectrum and two bulk EPs at $k=\pm \arccos{\frac{u^{2}-v^{2}-w^{2}}{2vw}}$ in Fig.~\ref{ptssheta}(a).  Next, without loss of generality, we choose a typical point $(w,u,v)=(0.7,0.5,0.8)$ in the $\mathcal{PT}$-broken phase to demonstrate $\eta$ and its derivative $\frac{d\eta}{dk}$ with the wave vector $k$ varying.  As shown in Fig.~\ref{ptssheta}(b-c), we find two discontinuity points of the derivative  $\frac{d\eta}{dk}$ located at the EPs, which is satisfied with the correspondence between the discontinuity of $\partial \eta$ and  EPs of bulk states. In conclusion, based on our numerical results and theoretical analysis, we propose that bulk EPs cause the discontinuity of the derivative of the quantity $\eta$, which is entirely different with edge EPs.

{\color{blue}\emph{Concluding remark.}}--- To measure non-unitarity  of non-Hermitian systems, we have defined a novel variant of Petermann factor $\eta$ which take on values within the interval $[0,1]$. As an efficient and powerful indicator of non-unitarity, the discontinuity of the quantity $\eta$ helps us identify rich physics  in non-Hermitian quantum systems.

In the context of the non-Hermitian lattice systems with EPs, the Hamiltonian matrix is classified as a defective matrix due to its lack of a complete basis of eigenvectors. Meanwhile, the numerical algorithm for diagonalizing such matrix is not convergence\cite{demmel1990nearest}. Therefore, it is challenging to directly identify the existence of EPs. Our introduced quantity $\eta$ provides alternative route to the features of bulk and edge EPs, e.g., by computing the behavior of $\eta$ in the parameter space and searching discontinuity. In conclusion, we report the introduction of $\eta$ and show its efficiency and usefulness in characterizing non-Hermitian physics.
For more concrete applications and a systematic analytic theory about $\eta$ (e.g., physics of the derivative of $\eta$ of all-th orders, and relation to entanglement \cite{chenEntanglement2021,chenQuantum2022,leePositionmomentum2014,leeFreefermion2015}), we leave them for future work.

 {\color{blue}\emph{Acknowledgements.}}---  This work was supported by NSFC Grant  No.~12074438, Guangdong Basic and Applied Basic Research Foundation under Grant No.~2020B1515120100, and the Open Project of Guangdong   Provincial Key Laboratory of Magnetoelectric Physics and Devices under Grant No.~2022B1212010008.

\providecommand{\noopsort}[1]{}\providecommand{\singleletter}[1]{#1}%

\appendix

\section*{Supplemental materials}

\section{Part-A: The effective bulk Hamiltonian of non-Hermitian SSH model-I~\eqref{ssh_model} with open boundary condition}
In this part, we discuss the effective bulk Hamiltonian of non-Hermitian SSH model-I~\eqref{ssh_model} with open boundary condition to study the properties of transition point. Based on Ref.~\cite{yaoEdge2018}, we do a similarity transition for the Hamiltonian matrix $H$ of the model-I~\eqref{ssh_model} with $N$ unit cells as
\begin{equation}
  \overline{H}= S^{-1} H S,
\end{equation}
where $S$ is a diagonal matrix whose diagonal elements are $\{1, 1, r, r, \cdots, r^{N-1}, r^{N-1} \}$. When take $r=\sqrt{(t_{2}-g)/(t_{2}+g)}$ and set $t_{2}=1,g=0.1$, $\overline{H}$ becomes the standard SSH model with intracell hopping  $t_{1}=\overline{t}_{1}$ and intercell hopping $\overline{t}_{2}=\sqrt{(t_{2}-g)(t_{2}+g)}$. Furthermore, the effective bulk Hamiltonian of the model-I~\eqref{ssh_model} with open boundary condition in $k$-space is written as
\begin{equation}
  \overline{H}(k) = (t_{1} + \overline{t}_{2}\cos k)\sigma_{x} + \overline{t}_{2} \sin k \sigma_{y}.
\end{equation}
From this expression, the transition points of the model-I~\eqref{ssh_model} are $t_{1}= \sqrt{t_{2}^{2}-g^{2}}$, which is consistent with our numerical results and the discussion in Ref.~\cite{chengCompetition2022}. Furthermore, the bulk properties of $H$ is determined by the effective bulk Hamiltonian $\overline{H}$. Therefore, the effective bulk states of $H$ can be constructed by using $\ket{\overline{\psi}_{n}}= e^{-i k n}\ket{\psi(k)}$, where $\ket{\psi(k)}$ is the eigenstates of $\overline{H}(k)$. Specifically, the right and left bulk effective states of Hamiltonian matrix $H$ are represented as $\ket{\psi^{R}_{n}}= S\ket{\overline{\psi}_{n}} = r^{n-1} e^{-i k n}\ket{\psi(k)}$ and $\bra{\psi^{L}_{n}} = \bra{\overline{\psi}_{n}}S^{-1} = \bra{\psi(k)}(1/r)^{n-1} e^{i k n} $. Then, based on above discussion, when $t_{2}=1,g=0.1$, we obtain that the inner product of left and right bulk states at transition point equal to $\braket{\psi^{L}_{n}|\psi^{R}_{m}} = \delta_{n,m}$. Due to the completeness of bi-orthogonal relation, we find that the bulk states at the transition point of the model-I~\eqref{ssh_model} do not have EPs.

\begin{figure}
        \centering
        \includegraphics[width=8cm]{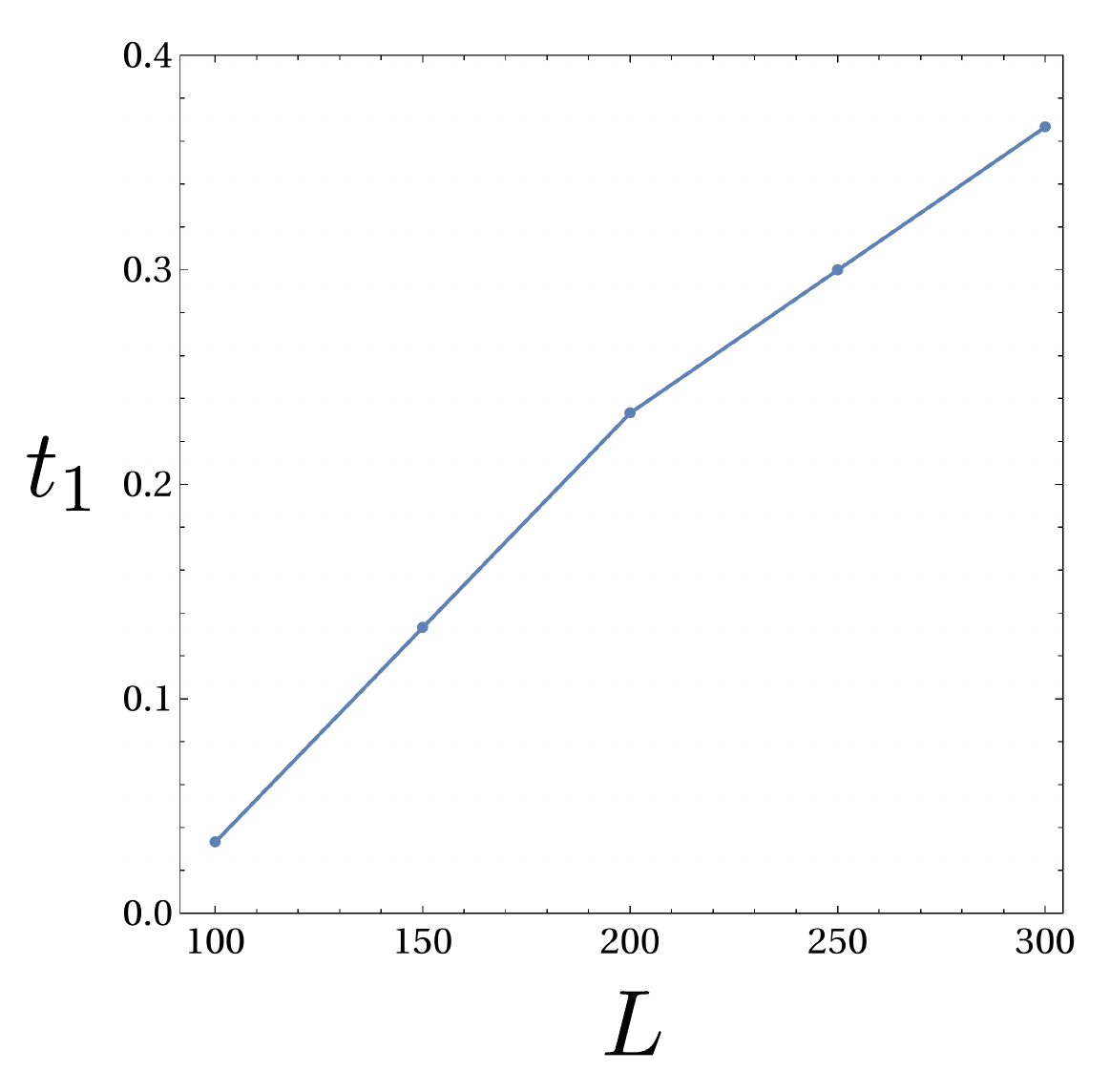}
        \caption{ The location of discontinuity point of $\eta$ with the size of the model-I increasing. Here $g=0.1$, $t_{2}=1$.}
        \label{fini_size}
\end{figure}

\section{Part-B: The analysis for the edge states in the non-Hermitian SSH model-I~\eqref{ssh_model}.}
In this part, we analytically obtain the edge states of the non-Hermitian SSH model-I~\eqref{ssh_model} with $N$ unit cells to give an explain about the edge state transition.  Based on Eq.~\eqref{ssh_model}, and we assume the edge states have the form $\ket{\psi_{\text{edge}}}=(\phi_{1,A},\phi_{1,B},\dots, \phi_{n,A},\phi_{n,B}, \dots,\phi_{N,A},\phi_{N,B})$, where $\phi_{n,A(B)}$ represents the wavefunction on the $A(B)$ sublattice in the $n$th unit cell. Then the eigenequation $H\ket{\psi_{\text{edge}}}=E_{\text{edge}}\ket{\psi_{\text{edge}}}$ can be concretely written as
\begin{equation}
  \begin{split}
    t_{1}\phi_{n,A} + (t_{2}-g)\phi_{n+1,A} &= E_{\text{edge}} \phi_{n,B}, n=1,2,\dots,N-1, \\
    (t_{2}+g) \phi_{n-1,B} + t_{1}\phi_{n,B} &= E_{\text{edge}}\phi_{n,A}, n=2,3,\dots,N. \\
  \end{split}
\end{equation}
Furthermore, the boundary conditions are represented as
\begin{equation}
  \begin{split}
    t_{1}\phi_{1,B} =E_{\text{edge}}\phi_{1,A}, \\
    t_{1}\phi_{N,A} = E_{\text{edge}} \phi_{N,B}.
  \end{split}
\end{equation}
When consider the thermodynamic limit (i.e., $N\rightarrow\infty$) and zero-edge states ($E_{\text{edge}}=0$), we obtain the zero-edge states written as
\begin{equation}
  \begin{split}
    \phi_{n,A} = (-\frac{t_{1}}{t_{2}-g})^{n-1}\phi_{1,A}, \\
    \phi_{1,B} = (-\frac{t_{1}}{t_{2}+g})^{n-1}\phi_{n,B}. \\
  \end{split}
\end{equation}
Meanwhile, the boundary conditions become as $\phi_{1,B}=\phi_{N,A}=0$. Without loss of generality, set $t_{2}=1, g=0.1$ and $t_{1}>0$, when $t_{1}<t_{2}-g$, then the wavefunctions $\phi_{n,A}$ and $\phi_{n,B}$ are respectively localized at the left and right endpoints of the 1D chain. Consequently, the two wavefunctions are orthogonal and have no contribution to the quantity $\eta$. When $t_{1}>t_{2}-g$,  to satisfy the boundary condition $\phi_{N,A}=0$, then $\phi_{N,A}=\phi_{1,A}=\phi_{n,A}=0$. Therefore, in this case, the two edge states merge into one edge states, and simultaneously localized at the right endpoint of the 1D chain. For this reason, the edge states of this case have the contribution to the quantity $\eta$. Therefore, with the parameter $t_{1}$ increasing from $t_{1}<t_{2}-g$ to $t_{1}>t_{2}-g$, the edge states of the model-I~\eqref{ssh_model} have edge state transition, and would cause the appearance of the discontinuity point of $\eta$ located at $t_{1}=t_{2}-g$. However, as shown in Fig.~\ref{sshmodel}(b), the discontinuity point does not appear at $t_{1}=t_{2}-g$, we propose the discrepancy of discontinuity point originates from the finite-size effect. As shown in Fig.~\ref{fini_size}, when we increase the size of model-I~\eqref{ssh_model}, we find that the discontinuity point of $\eta$ becomes more and more close to $t_{1}=t_{2}-g$, which is satisfied with our analysis.

\begin{figure}
        \centering
        \includegraphics[width=8.9cm]{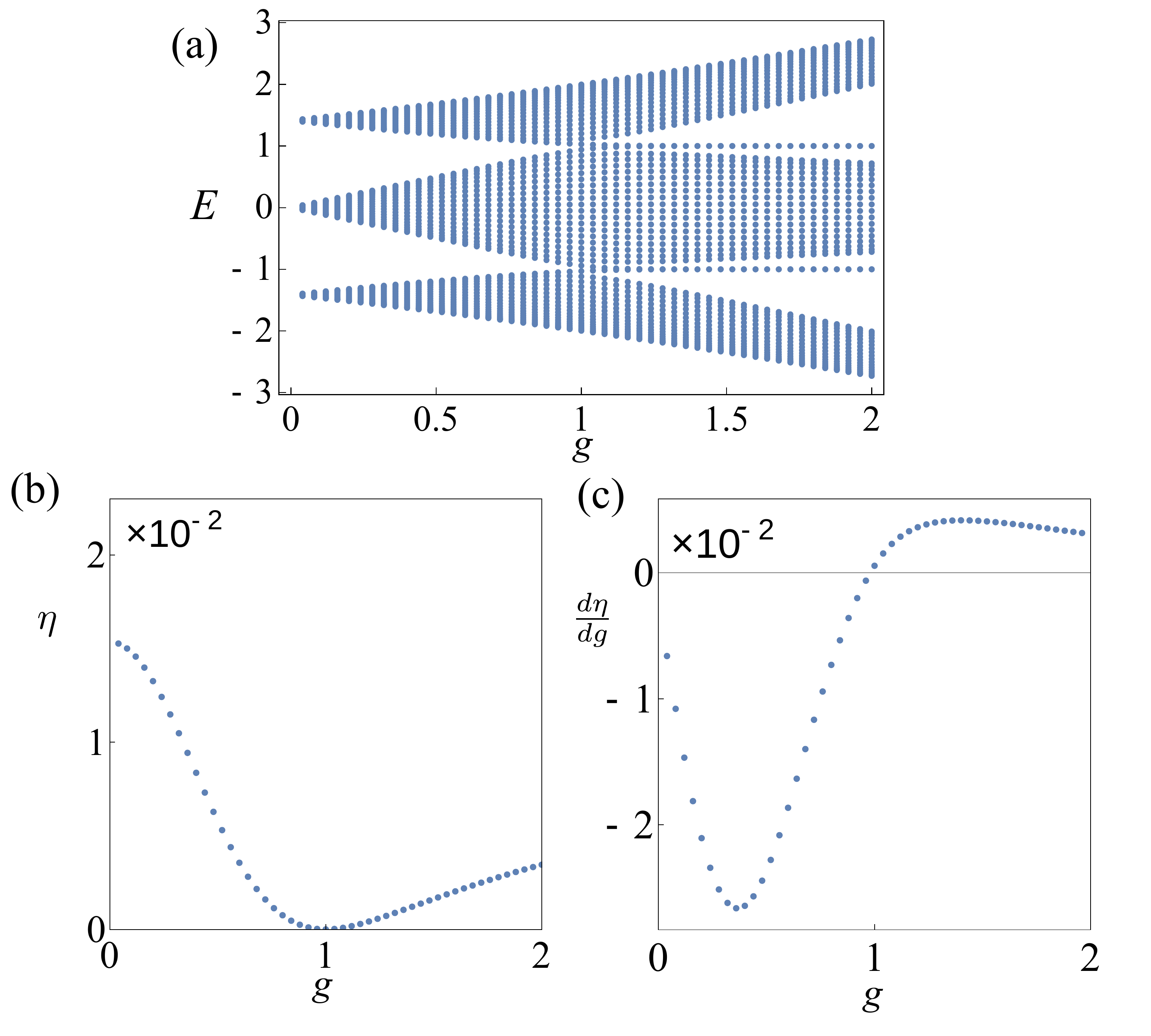}
        \caption{(a)Energy spectrum of the model~\eqref{zak_model} with open boundary condition as a function of the parameter $g$. (b) and (c) respectively show the quantity $\eta$ and its derivative as a function of parameter $g$.}
        \label{lioumodel}
\end{figure}

\section{Part-C: The behavior of quantity $\eta$ in a model without EPs. }
 To further investigate the connection between the discontinuity of $\partial\eta$ and the existence of EPs in bulk states, let us
   consider a $1$D non-Hermitian model~\cite{longNonHermitian2022} with topological phase transition at Hermitian point and absence of EPs:
\begin{equation}
    \begin{aligned}
    \textbf{Model-V} ~~ H = t_0 \sum_n &( \frac{1}{g}b_n^\dagger a_n +g a_n^\dagger b_n +\frac{1}{g}b_n^\dagger c_n +
    \\ &g c_n^\dagger b_n + g a_{n+1}^\dagger c_n + g c_n^\dagger a_{n+1}),
\end{aligned}
\label{zak_model}
\end{equation}
where $a_n(a_n^\dagger)$, $b_n(b_n^\dagger)$ and $c_n(c_n^\dagger)$ respectively denote the annihilation(creation) operator of spinless fermions at   sublattice A, B and C in the $n$th unit cell.  This model always has real energy spectrum without $\mathcal{PT}$ symmetry. When the parameter $g <1 $ ($g >1$), the system is trivial~(topological) phase. As discussed in Ref.~\cite{longNonHermitian2022}, the system with non-trivial Zak phase in the parameter range $g>1$ has topological edge states as shown in Fig.~\ref{lioumodel}(a).

Keeping the critical point $g=1$ in mind, we study the value and derivative of  $\eta$ as the functions of the ``non-Hermiticity inducer'' $g$ in the model-V~\eqref{zak_model} as shown in Fig.~\ref{lioumodel}(b) and (c). We find that $\eta$ reaches its minimum  exactly at  $g=1$ where the derivative vanishes and the model recovers Hermiticity. Apparently, the derivative of $\eta$ is always continuous in this model, which is different from   that in the models-III, IV~\eqref{quasi_model}. By careful analysis, we find  this difference originates from the model construction~\cite{longNonHermitian2022} of using the regular Sturm-Liouville theory~\cite{zettl2012sturm}( more discussion see Part-D). This theory guarantees the complete basis of the model-V~\eqref{zak_model}, so EP is absent in this system. Therefore, without EP, the quantity $\eta$ and its defective in this model-V~\eqref{zak_model} would not have discontinuity point.

\section{Part-D: The property of a class of special models and Sturm-Liouville theory}\label{app_liou}
In this part, we discuss the properties of the model-V \eqref{zak_model}. This model is constructed from the equation given as:
\begin{equation}
    H_{0}\Psi_{n} = E_{n} M \Psi_{n},
    \label{stu_liou}
\end{equation}
where $H_{0}$ is a Hermitian matrix, $M$ is a real diagonal matrix with diagonal element $M_{ii}>0$. This equation is discussed in the regular Sturm-Liouville theory~\cite{zettl2012sturm}. Meanwhile, as discussed in Ref.~\cite{longNonHermitian2022}, the Hamiltonian matrix of the model-V~\eqref{zak_model} is represented as $M^{-1} H_{0}$ and non-Hermitian.

Let us review some properties of Eq.\eqref{stu_liou} to demonstrate the model in Ref.~\cite{longNonHermitian2022} having real energy spectrum. Due to the property of the regular Sturm-Liouville theory~\cite{zettl2012sturm}, $0$ is not the eigenvalue of $H_{0}$. Then, Eq.~\eqref{stu_liou} can transformed to
\begin{equation}
    \lambda_{n}M^{\frac{1}{2}}\Psi_{n} = (M^{\frac{1}{2}} K M^{\frac{1}{2}}) M^{\frac{1}{2}}\Psi_{n},
    \label{stu_liou_tran}
\end{equation}
where $K$ is the inverse of $H_{0}$, and $\lambda_{n} = E_{n}^{-1}$. Furthermore, the matrix $K^{'} = M^{\frac{1}{2}} K M^{\frac{1}{2}}$ is Hermitian, which is proven by:
\begin{equation}
    (M^{\frac{1}{2}} K M^{\frac{1}{2}})^{\dagger} = (M^{\frac{1}{2}})^{\dagger} K^{\dagger}(M^{\frac{1}{2}})^{\dagger} = M^{\frac{1}{2}} KM^{\frac{1}{2}}.
\end{equation}
This proof is based on the Hermitian of the matrices $K$ and $M^{\frac{1}{2}}$. Therefore, from Eq.~\eqref{stu_liou_tran}, we find the eigenvalues $\lambda_{n}$ are real, and the eigenvalues $E_{n}$ of $M^{-1}H_{0}$ are real. Meanwhile, we find the eigenvectors $\Psi^{'}_{n} = M^{\frac{1}{2}}\Psi_{n}$ are a set of orthogonal and complete basis satisfying the relation
\begin{equation}
    \braket{\Psi^{'}_{n}|\Psi^{'}_{m}} = \bra{\Psi_{n}}M\ket{\Psi_{m}} = \delta_{n,m},
    \label{vect}
\end{equation}
where $\delta_{n,m}$ is the Kronecker delta. Based on the relation~\eqref{vect}, the eigenvectors $\Psi_{n}$ of the matrix $M^{-1} H_{0}$ are always complete. For a non-Hermitian system with EPs, the basis of eigenvectors is not complete. Then, EP is absent in the model-V~\eqref{zak_model}.

\end{document}